\documentclass[aps,prd,showpacs,twocolumn,groupedaddress]{revtex4}
\usepackage{graphicx}
\usepackage{amsfonts}
\usepackage{amssymb}
\usepackage{amsmath}
\def\lsim {~^{<~}_{\sim~}}

\usepackage{ulem}
\usepackage{color}

\begin{document}
\title{Gluon Propagators
in Maximally Abelian Gauge in SU(3) Lattice QCD}

\author{Shinya~Gongyo}
  \email{gongyo@ruby.scphys.kyoto-u.ac.jp}
  \affiliation{Department of Physics, Graduate School of Science,
  Kyoto University, \\
  Kitashirakawa-oiwake, Sakyo, Kyoto 606-8502, Japan}
\author{Hideo~Suganuma}
  \email{suganuma@ruby.scphys.kyoto-u.ac.jp}
  \affiliation{Department of Physics, Graduate School of Science,
  Kyoto University, \\
  Kitashirakawa-oiwake, Sakyo, Kyoto 606-8502, Japan}

\date{\today}
\begin{abstract}
In SU(3) lattice QCD, we study diagonal and off-diagonal gluon propagators in the maximally Abelian (MA) gauge with U(1)$_3 \times$U(1)$_8$ Landau gauge fixing. These propagators are studied both in the coordinate space and in the momentum space. The Monte Carlo simulation is performed on $16^4$ at $\beta$=6.0 and $32^4$ at $\beta$=5.8 and 6.0 at the quenched level. In the four-dimensional Euclidean space-time, the effective mass of diagonal gluons is estimated as $M_{\mathrm{diag}} \simeq 0.3\mathrm{GeV}$ and that of off-diagonal gluons as $M_{\mathrm{off}} \simeq 1\mathrm{GeV}$ in the region of $r =0.4-1.0$fm.   
In the momentum space, the effective mass of diagonal gluons is estimated as $M_{\mathrm{diag}} \simeq 0.3\mathrm{GeV}$ and that of off-diagonal gluons as $M_{\mathrm{off}} \simeq 1\mathrm{GeV}$ in the region of $p  <1.1$GeV. The off-diagonal gluon propagator is relatively suppressed in the infrared region and seems to be finite at zero momentum, while the diagonal gluon propagator is enhanced.
Furthermore, we also study the functional form of these propagators in momentum space. These propagators are well fitted by $Z/\left(p^2+m^2\right)^\nu$ with fit parameters, $Z, m$ and $\nu$ in the region of $p  <3.0$GeV. From the fit results and lattice calculations, all of the spectral functions of diagonal and off-diagonal gluons would have negative regions.
\end{abstract}
\maketitle

\section{Introduction}

Quantum chromodynamics (QCD) is the fundamental gauge theory of the strong interaction based on quarks and gluons. There are a variety of nonperturbative phenomena in low energy QCD such as color confinement and chiral symmetry breaking. These nonperturbative phenomena have been much studied \cite{C7980,R12,Cr11}, however, there are still open problems in QCD. In particular, it is thought that the confinement mechanism, how quarks and gluons are confined, is a considerably difficult problem.   

Nowadays, there are several confinement scenarios in various gauges. In the Landau gauge, confinement is related to the (deep) infrared behavior of the gluon propagator and the ghost propagator from the scenarios suggested by Kugo and Ojima, Gribov and Zwanziger \cite{KO79,G78,Z91}. These propagators have been much studied from the analytical frameworks such as Schwinger-Dyson equations \cite{AS01} and functional renormalization group equations \cite{P07} and lattice QCD calculations \cite{CM07,BIMS09,BMM09}.
 
On the other hand, the maximally abelian (MA) gauge has mainly been investigated from the viewpoint of the dual-superconductor picture \cite{GIS12,AS98,BC03,SST95,Ko11,SY90,IH99,Mi95,Wo95,KSW87,BWS91,SNW94,SAI02,HSA10,DGL04}, which is one of confinement scenarios suggested by Nambu, 't Hooft and Mandelstam \cite{N74}. This picture is based on the electromagnetic duality and the analogy with the one-dimensional squeezing of the magnetic flux in the type-II superconductor. In this picture, there occurs color magnetic monopole condensation, and then the color-electric flux between the quark and the antiquark is squeezed as a one-dimensional tube due to the dual Higgs mechanism.  From the viewpoint of the dual-superconductor picture in QCD, however, there are two assumptions of Abelian dominance \cite{tH81,EI82} and monopole condensation.
Here, Abelian dominance means that only the diagonal gluon component plays the dominant role for the nonperturbative QCD phenomena like confinement.

The various lattice QCD Monte Carlo simulations show that the MA gauge fixing seems to support these assumptions \cite{GIS12,AS98,BC03, SY90, IH99,Mi95, Wo95, KSW87, BWS91, SNW94,SAI02}. In fact, the diagonal gluons seem to be significant to the infrared QCD physics, which is called ``infrared Abelian dominance". Thus the difference between the diagonal gluon propagator and the off-diagonal gluon propagator in the infrared region seems to be significant.  

From this viewpoint, the gluon propagators in the MA gauge have been investigated in SU(2) lattice simulations \cite{AS98, BC03,Cu01} and in the SU(3) lattice simulation \cite{GIS12}. From these lattice simulations the off-diagonal gluons do not propagate in the infrared region due to the large effective mass $M_{\rm off} \simeq 1{\rm GeV}$, while the diagonal gluon widely propagates. In addition, the study of Schwinger-Dyson equations also supports the infrared Abelian dominance from the behavior of the scaling solutions, i.e., as momentum goes to zero, the diagonal gluon propagator 
is divergent and the off-diagonal gluon propagator is vanishing \cite{HSA10}. 

 The aim of this paper is to investigate the gluon propagators in the MA gauge of SU(3) lattice gauge theory. In Sec.\ref{2}, we briefly summarize the definition of MA gauge with U(1)$_3\times$U(1)$_8$ and some properties of lattice gluon propagators in this gauge. In Sec.\ref{3}, we show the method of estimating the effective mass in coordinate space and in momentum space. We study the gluon propagators in coordinate space in Sec.\ref{4}, and their functional form in momentum space in Sec.\ref{5}. Section~\ref{6} is devoted to the summary.

\section{SU(3) Formalism and gluon propagators in MA gauge with U(1)$_3\times$U(1)$_8$ Landau gauge}
\label{2}
Using SU(3) lattice QCD, we calculate the gluon propagators in the MA gauge with the U(1)$_3\times$U(1)$_8$ Landau gauge fixing. In the MA gauge, SU(3) gauge symmetry is partially fixed, and only U(1)$_3\times$U(1)$_8$ Abelian gauge symmetry remains. Accordingly, while diagonal gluons $A_\mu ^a (x) (a=3,8)$ behave as Abelian gauge fields, off-diagonal gluons $A_\mu ^a (x) (a\neq 3,8)$ behave as U(1)$_3\times$U(1)$_8$ charged matter fields. In the MA gauge, to investigate the gluon propagators, we use the gluon fields extracted directly from the link-variables \cite{GIS12,FN04}.
Here, we analytically extract gluon fields $A_\mu (x)$ from the link-variables $U_\mu(x) =e^{iag A_\mu (x)}$ with the lattice spacing $a$ and the gauge coupling $g$ as follows:
\begin{eqnarray}
A_\mu (x) = \frac{1}{iag}
\Omega  ^\dagger _\mu (x){\rm Ln} U_\mu ^d (x) \Omega _\mu (x),
\label{gluon definition}
\end{eqnarray}
where ${\rm Ln}$ is the natural logarithm
defined on complex numbers, $U^d_\mu (x)$ the diagonalized unitary matrix of $U_\mu (x)$, and $\Omega _\mu (x)$ the diagonalization unitary matrix,
\begin{eqnarray}
 U^d_\mu (x) = \Omega _\mu (x) U_\mu (x)\Omega^\dagger  _\mu (x).
\end{eqnarray}
We can obtain $U^d_\mu (x)$ and $\Omega _\mu (x)$ analytically by solving a cubic equation with real coefficients using Cardano's method \cite{GIS12}.

The MA gauge fixing is performed by the maximization of 
\begin{eqnarray}
R_{\rm MA}
	&\equiv&
	 \sum_x \sum^4_{\mu =1} 
	{\rm tr} \left[ U_\mu (x) \vec{H} U_\mu^{\dagger}(x) \vec{H}  \right],
						\label{eqn:Rma}
\end{eqnarray}
where $\vec{H} =(T_3,T_8)$ is the Cartan generator. 
In continuum limit, $agA_\mu(x) \rightarrow 0$, this gauge fixing is given by the minimization of 
\begin{eqnarray}
\sum_{a\neq 3.8}\int d^4x A_\mu ^a (x)A_\mu ^a(x).
\end{eqnarray}
Thus, this gauge fixing corresponds to minimizing the off-diagonal gluons under the gauge transformation. 

In this gauge fixing, there remains U(1)$_3\times$U(1)$_8$ gauge symmetry. In order to study the gluon propagators, we fix the residual gauge. After the Cartan decomposition for the SU(3) link-variables as $U_\mu (x) \equiv M_\mu (x) u_\mu (x)$ with $u_\mu(x) \equiv e^{i\left\{\theta^3(x)T ^3 + \theta ^8 (x) T ^8 \right\}}\in$U(1)$_3\times$U(1)$_8$ and $M_\mu (x)= e^{i\sum_{a\ne 3,8} \theta ^a (x)T^a} \in$SU(3)$/$U(1)$_3\times$U(1)$_8$, the residual gauge fixing is performed 
by the maximization of 
\begin{eqnarray}
R_{\rm U(1)L}\equiv \sum_x \sum_{\mu=1}^4 {\rm Re}\ {\rm tr}[u_\mu(x)].
											\label{eqn:U1Landau}
\end{eqnarray}
In our calculation, this procedure corresponds to ``minimal" MA gauge, 
and a random Gribov copy is taken in the Gribov region \cite{DGL04}. 
After gauge fixing completely, we study the diagonal (Abelian) gluon propagator as
\begin{eqnarray}
G_{\mu\nu}^{\rm diag}(x-y) \equiv \frac{1}{2} \sum_{a= 3,8} \left< A_\mu^a(x)A_\nu^a(y)\right>,
				\label{eqn:AAf002}
\end{eqnarray}
and the off-diagonal gluon propagator as
\begin{eqnarray}
G_{\mu\nu}^{\rm off}(x-y) \equiv
\frac{1}{6} \sum_{a\neq 3,8} \left< A_\mu^a(x)A_\nu^a(y)\right>.
				\label{eqn:AAf003}
\end{eqnarray}
Note that $G_{\mu\mu}^{\rm diag}(x-y)$ and $G_{\mu\mu}^{\rm off}(x-y)$  are expressed as the function of the four-dimensional Euclidean distance $r\equiv \sqrt{(x_\mu -y_\mu)^2}$. 
When we consider the renormalization, 
these propagators are multiplied by an $x$-independent constant, 
according to the renormalized gluon fields 
obtained by multiplying a constant renormalization factor.

Furthermore, on the periodic lattice of $L_1 \times L_2\times L_3\times L_4$, we consider the momentum-space gluon propagator, which is defined as
\begin{eqnarray}
G_{\mu\nu}^{\rm diag}(p) \equiv \frac{1}{2} \sum_{a= 3,8} \left< \tilde{A}_\mu^a(\tilde{p}) \tilde{A}_\nu^a(-\tilde{p})\right>,   \label{G_p^diag} \\
G_{\mu\nu}^{\rm off}(p) \equiv
\frac{1}{6} \sum_{a\neq 3,8} \left< \tilde{A}_\mu^a(\tilde{p})\tilde{A}_\nu^a(-\tilde{p})\right>,
				\label{G_p^off}
\end{eqnarray}
where $\tilde{p}$ and $p$ are defined as
\begin{eqnarray}
\tilde{p}_\mu\equiv \frac{2\pi n_\mu}{aL_\mu},~ p_\mu \equiv \frac{2}{a}\sin \left( \frac{\tilde{p}_\mu a}{2} \right) ,
\end{eqnarray}
with $n_\mu = 0,1,2, \ldots ,L_\mu -1$, 
and $\tilde{A}^a_\mu (\tilde{p})$ is defined as
 \begin{eqnarray}
\tilde{A}^a_\mu (\tilde{p})  = \sum _x e^{-i\tilde{p}_\nu x_\nu -\frac{i}{2}\tilde{p}_\mu a} A^a_\mu (x).
\end{eqnarray}
Note that when the local Landau gauge fixing is satisfied as $\partial _\mu A_\mu ^a (x) = 0$, this condition is expressed by
 \begin{eqnarray}
 p_\mu \tilde{A}^a_\mu (\tilde{p}) =0 ~~~({\rm Landau}) \label{landauq}
 \end{eqnarray}
in momentum space. In continuum limit, $agA_\mu(x) \rightarrow 0$, the global gauge fixing condition to maximize $R_{\rm U(1)L}$ in Eq. (\ref{eqn:U1Landau}) reduces to the local U(1)$_3 \times $U(1)$_8$ Landau condition $\partial _\mu A_\mu ^a (x) = 0~(a=3,8)$. Therefore, Eq. (\ref{landauq}) is satisfied for the diagonal components in momentum space. Then, the diagonal gluon propagator has only the transverse component $T^{\rm diag}(p^2)$:
\begin{eqnarray}
G_{\mu\nu}^{\rm diag}(p) = \left(\delta _{\mu\nu} - \frac{p_\mu p_\nu}{p^2}\right) T^{\rm diag}(p^2).
\end{eqnarray} 

On the other hand, the off-diagonal gluon propagator consists of two components corresponding to longitudinal and transverse components. 
In the continuum limit, the local gauge fixing condition 
for the off-diagonal gluons is given by \cite{IH99} 
\begin{eqnarray}
[\vec H, [\hat D_\mu, [\hat D_\mu, \vec H]]]=0
\end{eqnarray}
with the covariant derivative $\hat D_\mu$, and this condition 
is different from the Landau gauge fixing (\ref{landauq}). 
Therefore, the off-diagonal gluon propagator has 
the longitudinal component even in the continuum limit. 
Then, the off-diagonal gluon propagator is expressed by 
\begin{eqnarray}
G_{\mu\nu}^{\rm off}(p) = \left(\delta _{\mu\nu} - \frac{p_\mu p_\nu}{p^2}\right) T^{\rm off}(p^2) + \frac{p_\mu p_\nu}{p^2}L^{\rm off}(p^2),
\end{eqnarray}
with the longitudinal component $L^{\rm off}(p^2)$ and the transverse component $T^{\rm off}(p^2)$. These two functions are derived from the off-diagonal gluon propagator as follows:
\begin{eqnarray}
L^{\rm off}(p^2) &=& \frac{p_\mu p_\nu}{p^2}G_{\mu\nu}^{\rm off}(p), \\
T^{\rm off}(p^2) &=& \frac{1}{3}\left( G_{\mu\mu}^{\rm off}(p) - \frac{p_\mu p_\nu}{p^2}G_{\mu\nu}^{\rm off}(p) \right).
\end{eqnarray}

In this way, we also study the transverse function of the diagonal propagator, and the transverse and longitudinal functions of the off-diagonal propagator in momentum space.

The Monte Carlo simulation is performed  with the standard plaquette action on the $16^4$ lattice with $\beta$ =6.0 and  on the $32^4$ lattice with $\beta$ =5.8 and 6.0 at the quenched level. The lattice spacings $a$ are determined so as to reproduce the string tension $\sigma \simeq 0.89 \mathrm{GeV/fm} $. At $\beta$=5.8 and 6.0, the lattice spacings $a$ are estimated as $a \simeq 0.186{\rm fm}$ and $0.152{\rm fm}$, respectively \cite{IS09}.
 All measurements are done every 500 sweeps after a thermalization of 10,000 sweeps using the pseudo heat-bath algorithm. We mainly use 50 configurations for the $16^4$ lattice and 20 configurations for the $32^4$ lattice at each $\beta$. The error is estimated with the jackknife analysis.

\section{Propagator in the Proca formalism}
\label{3}
Next, we investigate the effective gluon mass.
We start from the Lagrangian of 
the free massive vector field $A_\mu$ with the mass $M \ne 0$ 
in the Proca formalism,
\begin{eqnarray}
{\cal L}&=& \frac{1}{4}(\partial_\mu A_\nu - \partial _\nu A_\mu)^2
	+\frac{1}{2}M^2A_\mu A_\mu,   \label{eqn:Lag} 
\end{eqnarray}
in the Euclidean metric. The scalar combination of the propagator $G_{\mu\mu}(r;M)$ can be expressed 
with the modified Bessel function $K_1(z)$ as 
\begin{eqnarray}
{G}_{\mu\mu}(r;M) &=& \left< A_\mu(x) A_\mu(y) \right> \nonumber \\
	&=&\int\frac{d^4 k}{(2\pi)^4} e^{i k \cdot (x-y)}
	 \frac{1}{k^2+M^2}
 	\left( 4+\frac{k^2}{M^2}\right) \nonumber\\
	&=&
	3\int\frac{d^4k}{(2\pi)^4} e^{i k \cdot (x-y)}	\frac{1}{k^2+M^2}    
	 + \frac{1}{M^2}{\delta^4(x-y)} \nonumber \\
	&=&
	\frac{3}{4\pi^2}\frac{M}{r}K_1(Mr)+\frac{1}{M^2}\delta^4(x-y).
					\label{eqn:prp02}
\end{eqnarray}
In the infrared region with large $Mr$, 
Eq. (\ref{eqn:prp02}) reduces to 
\begin{eqnarray}
G_{\mu\mu}(r;M)
	&\simeq&
	\frac{3\sqrt{M}}{2(2\pi)^{\frac{3}{2}}} \frac{e^{-Mr}}{r^\frac{3}{2}}, 
						\label{eqn:prp03} 
\end{eqnarray}
using the asymptotic expansion, 
\begin{eqnarray}
K_1(z) \simeq
	\sqrt{\frac{\pi}{2z}} e^{-z} 
	\sum^\infty_{n=0}
	\frac{\Gamma(\frac{3}{2}+n)}{n!\Gamma(\frac{3}{2}-n)} \frac{1}{(2z)^n}, 
						\label{eqn:prp03A}
\end{eqnarray}
for large ${\rm Re}~z$. Here $\Gamma$ is the gamma function. Then, from the slope of the lattice QCD data of $\ln\left(r^{3/2}G_{\mu\mu}(r)\right)$, the effective gluon mass is estimated.

On the other hand, the propagator in momentum space $\tilde{G}_{\mu\nu}(p;M)$ can be expressed by
\begin{eqnarray}
\tilde{G}_{\mu\nu}(p;M) = \left( \delta_{\mu\nu} -\frac{p_\mu p_\nu}{p^2}\right) \frac{1}{p^2+M^2} + \frac{p_\mu p_\nu}{M^2}.
\end{eqnarray}
Therefore, the longitudinal component $L(p^2)$ and the transverse component $T(p^2)$ in the Proca-formalism propagator are regarded as
\begin{eqnarray}
 L(p^2) = \frac{1}{M^2},~ T(p^2) = \frac{1}{p^2 + M^2}. \label{LandT}
 \end{eqnarray}
Thus, by comparing the lattice QCD data of $L(p^2)$ and $T(p^2)$ with Eq. (\ref{LandT}), the effective gluon mass can be estimated.
\section{Analysis of gluon propagators in MA gauge in coordinate space}
\label{4}
\begin{figure}[h]
\begin{center}
\includegraphics[scale=0.4]{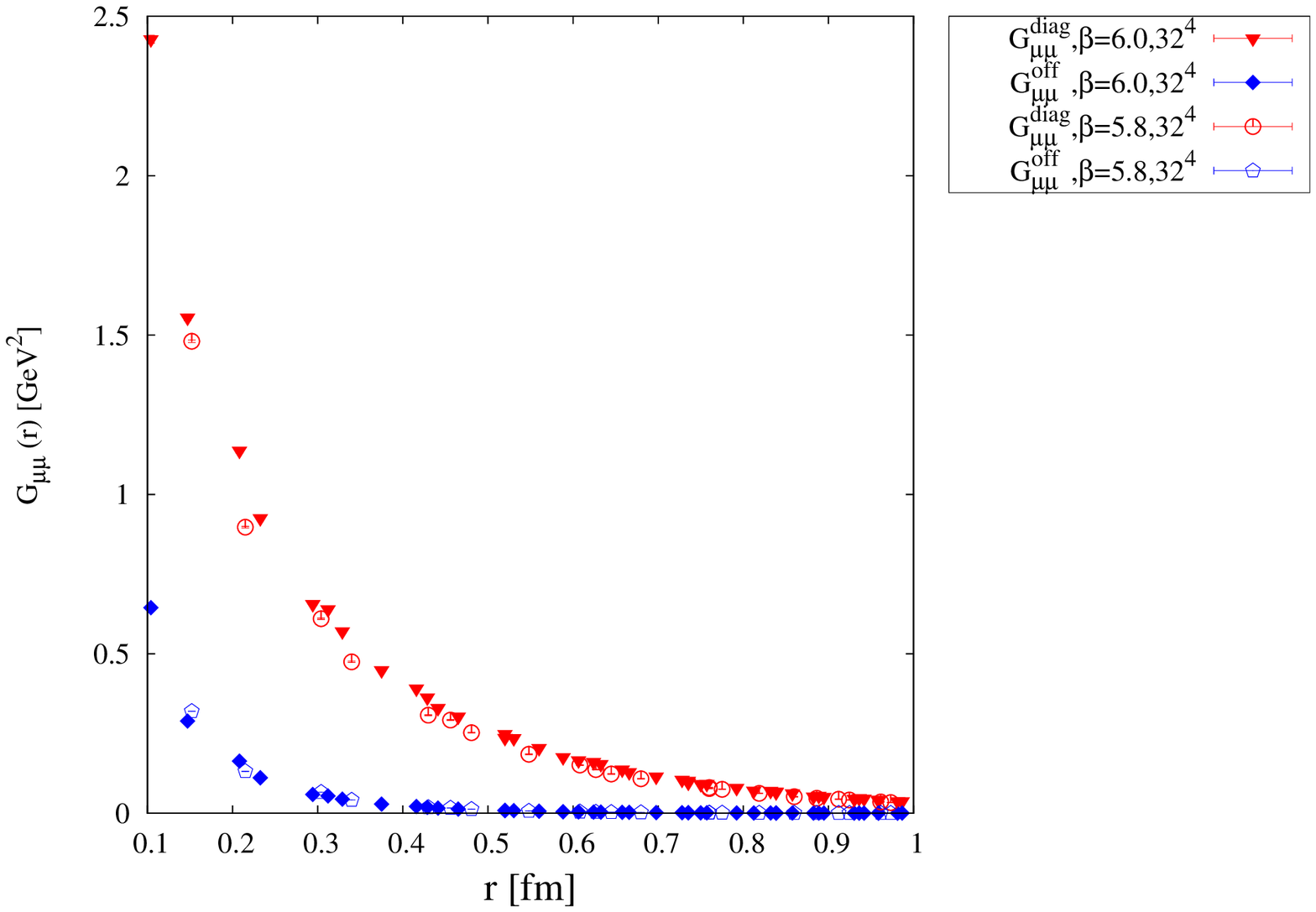}
\includegraphics[scale=0.4]{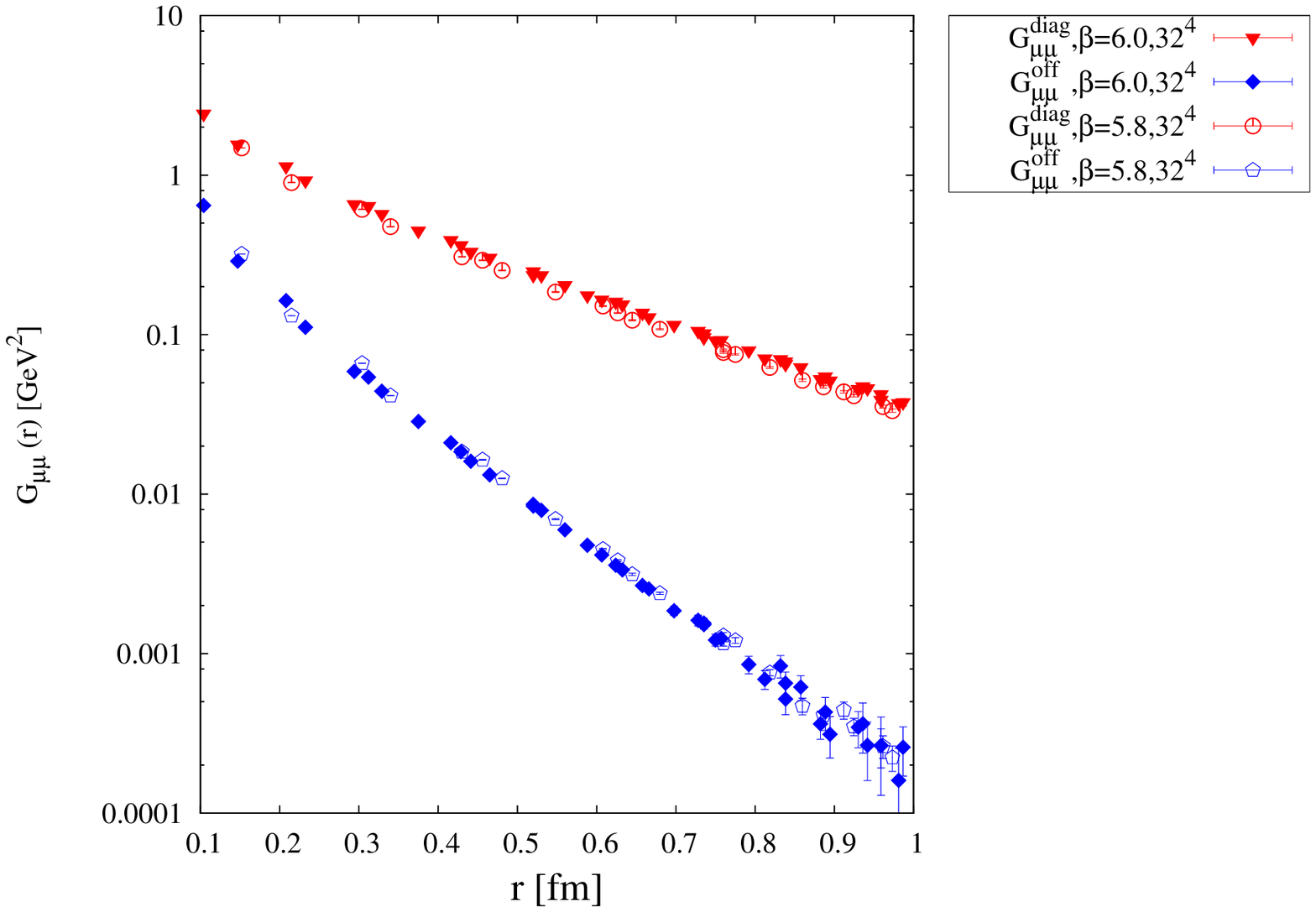}
\caption{
The SU(3) lattice QCD results of gluon propagators $G_{\mu\mu}^{\rm diag}(r)$ and $G_{\mu\mu}^{\rm off}(r)$ (top), and their logarithmic plots (bottom) as the function of $r\equiv \sqrt{(x_\mu -y_\mu)^2}$ in the MA gauge with the U(1)$_3\times$U(1)$_8$ Landau gauge fixing in the physical unit. The Monte Carlo simulation is performed on the $32^4$ lattice with $\beta$ = 5.8 and 6.0.
}
\label{Fig1a-b}
\end{center}
\end{figure}
We show in Fig.\ref{Fig1a-b} the lattice QCD result for the diagonal gluon propagator $G_{\mu\mu}^{\rm Abel}(r)$ and the off-diagonal gluon propagator $G_{\mu\mu}^{\rm off}(r)$ in the MA gauge with the U(1)$_3\times$U(1)$_8$ Landau gauge fixing on $32^4$ lattice with $\beta$ = 5.8 and 6.0. At the long distance, the off-diagonal gluon propagator is largely reduced, while the diagonal-gluon propagator takes a large value. The infrared abelian dominance is thus found in the MA gauge.

\begin{figure}[h]
\begin{center}
\includegraphics[scale=0.4]{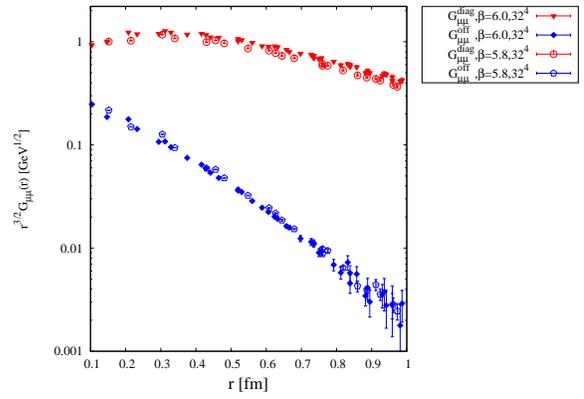}
\caption{
The logarithmic plot of $r^{3/2}G_{\mu\mu}^{\rm off} (r)$ and $r^{3/2}G_{\mu\mu}^{{\rm diag}} (r)$ as the function of the distance $r$ in the MA gauge with the U(1)$_3\times$U(1)$_8$ Landau gauge fixing, using the SU(3) lattice QCD with $32^4$ at $\beta$ = 5.8 and 6.0.
}
\label{Fig1}
\end{center}
\end{figure}
To evaluate the infrared abelian dominance quantitatively, we estimate the diagonal and off-diagonal gluon masses in coordinate space from the slope on the logarithmic plot of  $r^{3/2}G_{\mu\mu}^{\rm diag} (r)$ and $r^{3/2}G_{\mu\mu}^{\rm off} (r)$, respectively.
In Fig.\ref{Fig1}, we show the logarithmic plot of $r^{3/2}G_{\mu\mu}^{\rm off} (r)$ and $r^{3/2}G_{\mu\mu}^{\rm diag} (r)$ as the function of the distance $r$ in the MA gauge with the U(1)$_3\times$U(1)$_8$ Landau gauge fixing.
Note that the gluon-field renormalization does not affect the gluon mass 
estimate, since it gives only an overall constant factor for the propagator. 
We summarize in Table~I the effective off-diagonal gluon mass $M_{\rm eff}$ and the diagonal gluon mass $M_{\rm diag}$
obtained from the slope analysis in the range of $r=0.4-0.8{\rm fm}$ 
at $\beta$ =5.8 and 6.0 on $32^4$.
The off-diagonal gluons seem to have a large effective mass of 
$M_{\rm off} \simeq 1.1 {\rm GeV}$, while the diagonal gluon mass seems to be $M_{\rm diag} \simeq 0.3 {\rm GeV}$.
This means that only the diagonal gluons $A_\mu^3,A_\mu^8$ in the MA gauge 
propagate over the long distance and the infrared abelian dominance is found in the MA gauge. 
This result approximately coincides with our previous work \cite{GIS12}.
\begin{table}[h]
\caption{Summary table of conditions and results in SU(3) lattice QCD. 
The off-diagonal gluon mass $M_{\rm off}$ and the diagonal gluon mass $M_{\rm diag}$ are 
estimated from the slope analysis of $r^{3/2}G_{\mu\mu}^{\rm off} (r)$ and $r^{3/2}G_{\mu\mu}^{\rm diag} (r)$ for $r=0.4-1.0{\rm fm}$ at each $\beta$.
In the MA gauge, the off-diagonal gluon mass is estimated as $M_{\rm off} \simeq 1.1\mathrm{GeV}$, while the diagonal gluon mass estimated as $M_{\rm diag} \simeq 0.3\mathrm{GeV}$.
}
\begin{center} 
\begin{tabular}{ccccc}
\hline
\hline
  lattice size   & $\beta$     & $a[{\rm fm}]$ &   $M_{\rm off} [{\rm GeV}] $  &   $M_{\rm diag} [{\rm GeV}] $ \\
\hline
$32^4$                 &    ~~~5.8~~~  & 0.152 &  1.1  & 0.3\\
                 &    ~~~6.0~~~     &  0.104 & 1.1 & 0.3\\
\hline
\hline
\end{tabular}
\end{center} 
\end{table}

Here, these effective masses are to be considered to give 
the exponential damping of the correlation, and they are 
not simple pole masses. Actually, as will be shown in Sec.\ref{5}C, 
the functional form of the gluon propagator 
does not indicate a simple massive propagator with a definite pole mass.

Note that, because of $M_{\rm off} \gg M_{\rm diag}$, 
the diagonal-gluon propagation is dominant 
at the long distance as
\begin{eqnarray}
\frac{G_{\mu\mu}^{\rm off} (r)}{G_{\mu\mu}^{\rm diag} (r)} \sim \frac{e^{-M_{\rm off}r}}{e^{-M_{\rm diag}r }}\sim e^{-\left( M_{\rm off}-M_{\rm diag}  \right)r} \rightarrow 0,
\end{eqnarray}
and, accordingly, the diagonal-gluon contribution becomes 
decisively dominant in the infrared region.
In this way, the quantitative difference between the diagonal and the off-diagonal effective mass leads to the qualitative difference 
for the long-distance physics. 
This situation is similar to the pion contribution to the nuclear force. 
At the long distance, the nuclear force is well described by 
only one-pion exchange, 
and the contribution from other heavy mesons can be neglected.
In fact, the long-distance physics is dominated by light fields. 

\begin{figure}[h]
\begin{center}
\includegraphics[scale=0.4]{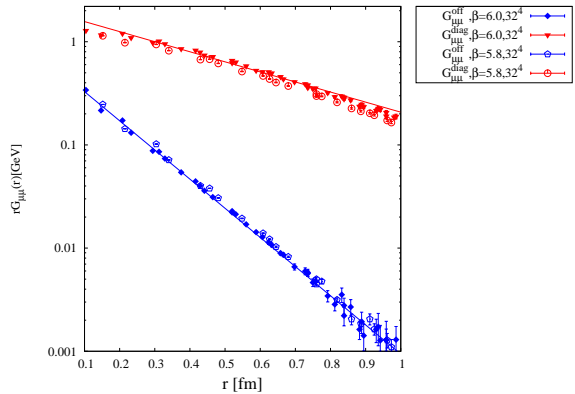}
\caption{
The logarithmic plot of 
$rG_{\mu\mu}^{\rm off} (r)$ and $rG_{\mu\mu}^{{\rm diag}} (r)$ 
as the function of the four-dimensional Euclidean distance $r$ 
in the MA gauge with the U(1)$_3\times$U(1)$_8$ Landau gauge fixing with $32^4$ at $\beta$=5.8 and 6.0. The solid line denotes the best-fit four-dimensional Yukawa function.
For $rG_{\mu\mu}^{\rm off} (r)$, 
the approximate linear correlation is found for Yukawa-fit mass parameter $m\simeq 1.3 {\rm GeV}$.
}
\label{Fig1d}
\end{center}
\end{figure}

Next, we also consider the functional form of diagonal and off-diagonal gluon propagators. In our previous work \cite{GIS12}, the off-diagonal gluon propagator is well described by the four-dimensional Euclidean Yukawa function as $ e^{-mr}/r$ with the mass parameter $m$ in the region of $r=0.1-0.8 {\rm fm}$. In Fig.\ref{Fig1d}, we show the logarithmic plot of $rG_{\mu\mu}^{\rm off} (r)$ and $rG_{\mu\mu}^{\rm diag} (r)$ as the function of the distance $r$ in the MA gauge with the U(1)$_3\times$U(1)$_8$ Landau gauge fixing. The logarithmic plot of $rG_{\mu\mu}^{\rm off} (r)$ seems to be linear. From the fit result by the four-dimensional Yukawa function in the region of $r=0.1-1.0 {\rm fm}$, the mass parameter $m$ lies in $1.3 {\rm GeV}$ for off-diagonal gluons.

On the other hand, from Fig.\ref{Fig1d}, the diagonal gluon propagator is not well described by the four-dimensional Yukawa function at least in short distance. In Sec.\ref{5}, we consider the functional form of the diagonal gluon propagator in momentum space.

We comment on the two types of gluon-field definitions on lattice, 
and their coincidence on the gluon propagator. 
In this paper, the gluon fields $A_\mu ^a (x)$ are extracted from 
$U_\mu (x) = e^{iag A_\mu (x)}$. 
On the other hand, with the Cartan decomposition of 
$U_\mu (x)=M_\mu(x)u_\mu (x)$, 
one can take another definition of the diagonal gluons extracted from 
$u_\mu (x)=e^{iag  \sum_{a = 3,8}{\cal A}_\mu ^a(x) T^a}$ 
and the off-diagonal gluons extracted from 
$M_\mu(x)=e^{iag  \sum_{a \neq 3,8}{\cal A}_\mu ^a(x) T^a}$. 
In the continuum limit, these two definitions of gluon fields coincide.
In our lattice calculation, even with these definitions of gluons, 
we obtain almost the same results for both 
diagonal and off-diagonal propagators. 
In Appendix \ref{ap1}, we also estimate the diagonal gluon mass 
in the Cartan-decomposition formalism. 

\section{Analysis of gluon propagators in MA gauge in momentum space}
\label{5}
\subsection{Gluon propagators in MA gauge in momentum space}
\label{5a}
\begin{figure}[h]
\begin{center}
\includegraphics[scale=0.5]{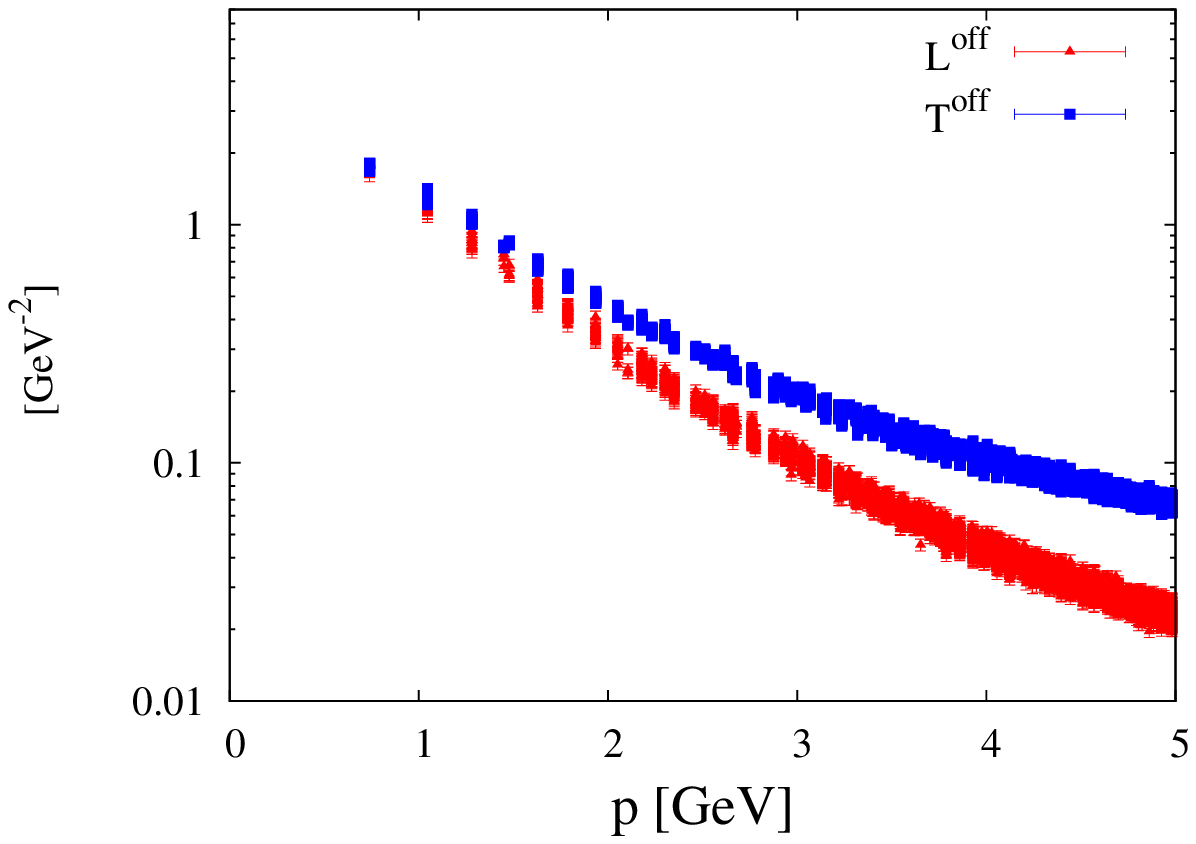}
\includegraphics[scale=0.5]{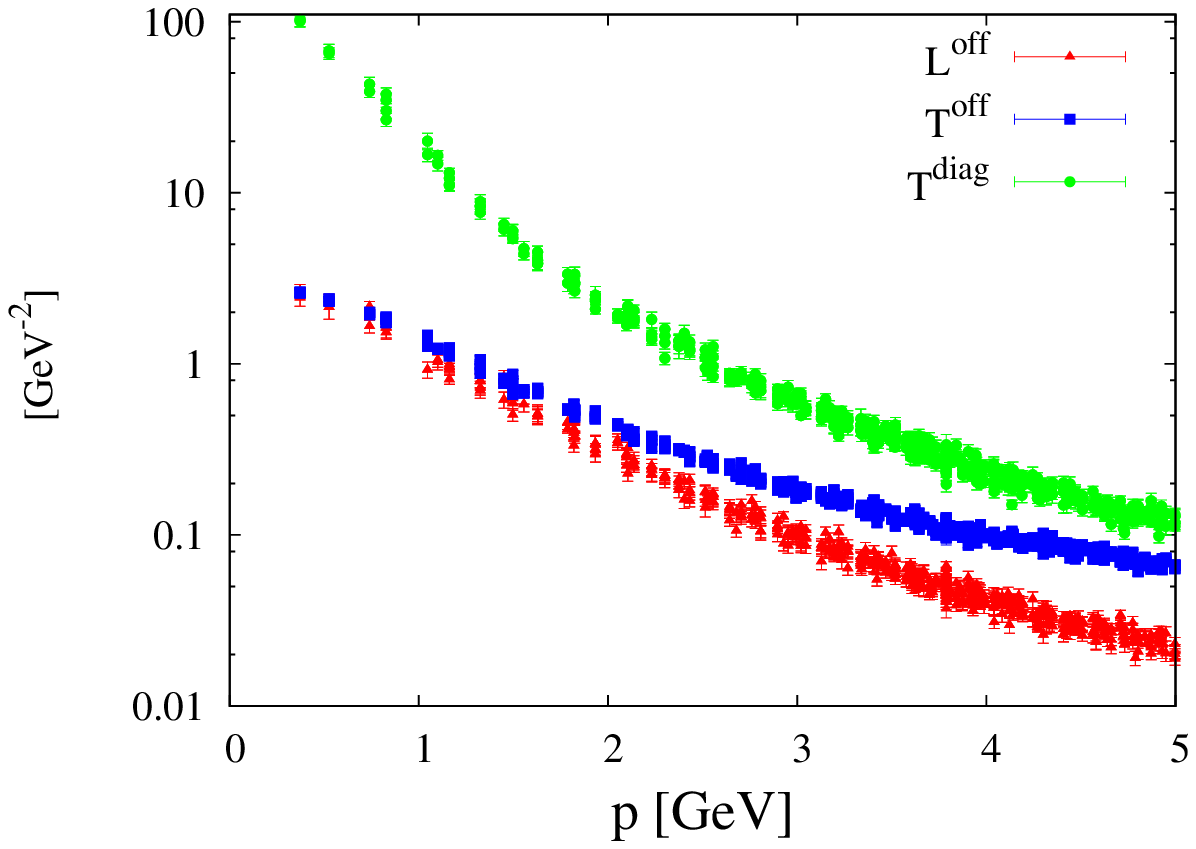}
\caption{
The logarithmic plot of $T^{\rm off} (p^2)$,$L^{\rm off} (p^2)$ and $T^{\rm diag} (p^2)$ as the function of the momentum $p\equiv (p_\mu p_\mu)^{1/2}$ with $16^4$ (top) and $32^4$ (bottom) at $\beta$=6.0. The diagonal gluon propagator is calculated only on $32^4$ due to large volume dependence. In infrared region, the diagonal gluon propagator is largely enhanced, while the longitudinal and transverse parts in the off-diagonal gluon propagator are relatively suppressed. The infrared Abelian dominance is found also in momentum space.}
\label{Fig2}
\end{center}
\end{figure}

In this section, we investigate the momentum-space propagators in the MA gauge in SU(3) lattice QCD. In Fig.\ref{Fig2}, we show the transverse and longitudinal components of off-diagonal propagator and the transverse component of diagonal propagator. Here, we calculate them with the two different lattice size, $16^4$ and $32^4$ at $\beta = 6.0$. As for the diagonal gluon propagator, we only adopt $32^4$, because its behavior largely depends on the lattice size \cite{GIS12}, while off-diagonal gluons have small volume-dependence. Note that, in MA gauge with U(1)$_3\times$ U(1)$_8$ Landau gauge, there is almost no the longitudinal diagonal component, which is checked in Appendix \ref{ap2}.   

In low momentum ($p\lsim 1{\rm GeV}$), the behavior of the longitudinal part is similar to that of the transverse part. From the infrared behavior, these parts would be finite at zero momentum, 
\begin{eqnarray}
L^{\rm off}(p^2) \simeq T^{\rm off}(p^2) \neq 0.
\end{eqnarray}
On the other hand, in high momentum, the longitudinal part is larger reduced than the transverse one. These behavior indicates that the Proca propagator is not suitable for the off-diagonal propagator as the functional form in the whole momentum region.

The diagonal part is larger enhanced in the infrared region than the off-diagonal parts and thus dominates the infrared nonperturbative phenomena. This suggests the infrared Abelian dominance. At zero momentum, we cannot determine from the infrared behavior whether the diagonal part is finite or not. As momentum goes to zero, the diagonal propagator seems to be still finite, however, for the definite argument in the deep infrared region, more careful analysis with a larger lattice size would be needed. 

\subsection{Estimation of gluon mass in MA gauge in momentum space}
\label{5b}
First, we estimate the effective mass of gluons by comparing the obtained gluon propagators with the massive vector-boson propagator in momentum space. The longitudinal and transverse components in the Proca formalism have the common factor, according to a constant renormalization factor $Z$ of the renormalized gluon fields,
\begin{eqnarray}
 L(p^2) = \frac{Z}{M^2},~ T(p^2) = \frac{Z}{p^2 + M^2}.
 \end{eqnarray}

As for the off-diagonal propagator, the transverse component is fitted by $\frac{Z}{p^2 + M^2}$ with the fit parameters, $Z$ and $M$, in the region of $p < p_{\rm max}=1.1{\rm GeV}$. Fixing the constant $Z$, we also fit the longitudinal component by $\frac{Z}{p^2 + M^2}$ with a fit parameter $M$ in the region of $p < p_{\rm max}$.

The diagonal transverse component is also fitted in a similar manner to the off-diagonal transverse one.

We summarize in Table~II the fit results in the region of $p<p_{\rm max}=1.1{\rm GeV}$. The off-diagonal gluon masses from the transverse and the longitudinal parts almost coincide as
\begin{eqnarray}
M_{\rm off}^{T} \simeq M_{\rm off}^{L} \simeq 1 {\rm GeV}.
\end{eqnarray} 
On the other hand, the diagonal gluon seems to behave with the effective mass $M_{\rm diag}^T \simeq 0.3 {\rm GeV}$ from the transverse part result. These results coincide with the coordinate results in Sec.\ref{4}.
\begin{table}[h]
\caption{Summary table of the gluon mass estimation in momentum space. 
The off-diagonal gluon mass $M_{\rm off}$ and the diagonal gluon mass $M_{\rm diag}$ are 
estimated from the fit analysis by the Proca propagator for $p<1.1{\rm GeV}$.
In the MA gauge, the off-diagonal gluon mass is estimated as $M_{\rm off} \simeq 1\mathrm{GeV}$, while the diagonal gluon mass estimated as $M_{\rm diag} \simeq 0.3\mathrm{GeV}$. These results coincide with the coordinate results.
}
\begin{center} 
\begin{tabular}{cccccc}
\hline
\hline
&  lattice size ~  & $p_{\rm max} [\rm GeV]$     &   $M [{\rm GeV}] $  & $Z$ &   $\chi ^2/N$ \\
\hline
$T^{\rm off}$&$32^4$&    ~~~1.1~~~  & 0.93(2) &  2.7(1) & 0.9\\
$L^{\rm off}$&$32^4$&    ~~~1.1~~~  & 1.00(2) &  2.7(fix) & 2.5\\
$T^{\rm diag}$&$32^4$&    ~~~1.1~~~ &  0.26(2)  & 22.0(8) & 1.7\\
\hline
\hline
\end{tabular}
\end{center} 
\end{table}

\subsection{Analysis of functional forms in MA gauge in momentum space}
As the fit region increases, gluon propagators cannot be described well with the massive vector boson propagator, $\frac{Z}{p^2+M^2}$. In fact, in the region of $p<3.0{\rm GeV}$, $\chi ^2 /N$ of the off-diagonal transverse part is larger than 3, and that of the diagonal transverse part is larger than 15. This result seems to be consistent with our previous work \cite{GIS12}. In coordinate space, the functional form of the off-diagonal gluon propagator has not be described by the propagator of the Proca formalism, but by the four-dimensional Euclidean Yukawa function $e^{-mr}/r$ with a mass parameter $m$. The Fourier transformation of this Yukawa function is expressed by
\begin{eqnarray}
 \int d^4x e^{ip\cdot x} \frac{e^{-mr}}{r} = \frac{4\pi^2}{(p^2 + m^2)^{3/2}}.
 \end{eqnarray}
 
 Thus, we fit the transverse and longitudinal part of the off-diagonal propagator and the transverse part of the diagonal propagator by
 \begin{eqnarray}
 \frac{Z}{(p^2 + m^2)^{3/2}}
 \end{eqnarray}
 with the fit parameters $Z$ and $m$ in the region of $p< p_{\rm max}= 3.0{\rm GeV}$.
 Note that $Z$ cannot be regarded as a renormalization constant, and, in general, the fit parameters $Z$ in the transverse part and the longitudinal part would differ.
\begin{figure}[h]
\begin{center}
\includegraphics[scale=0.5]{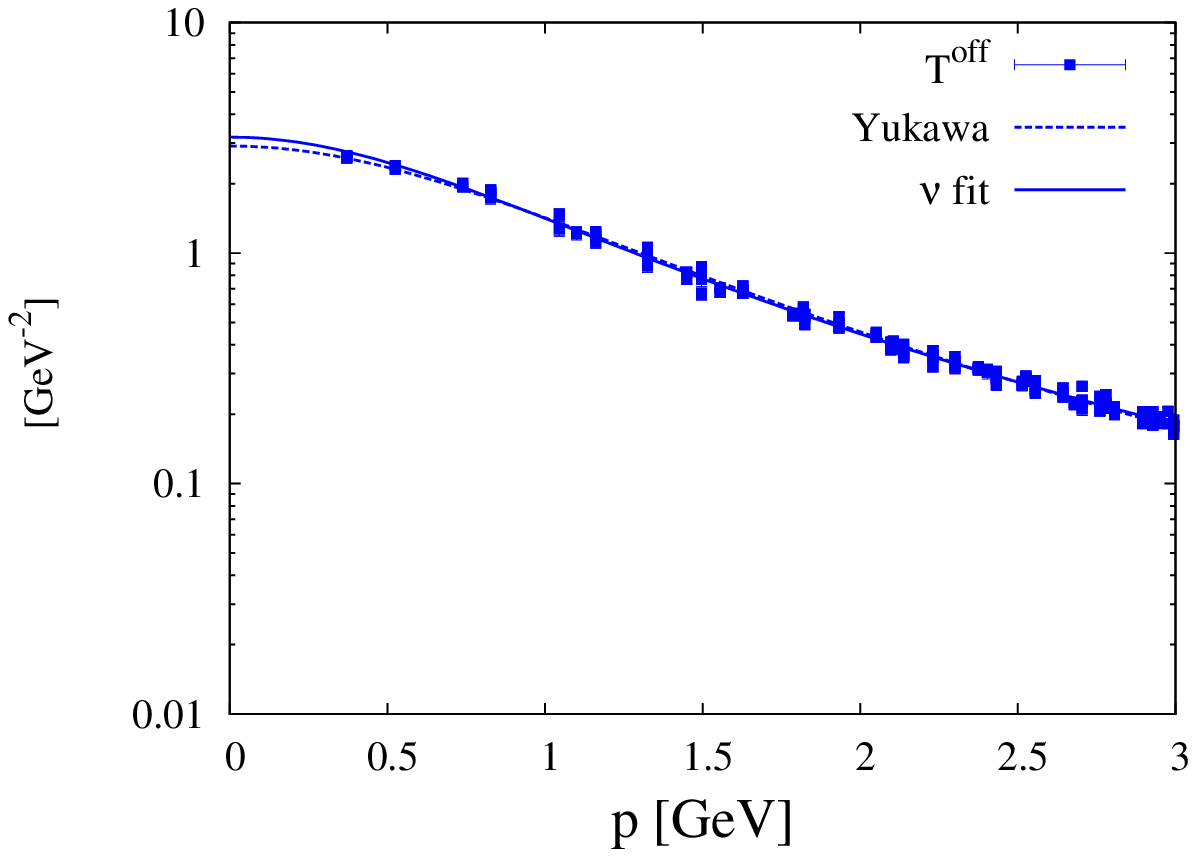}
\includegraphics[scale=0.5]{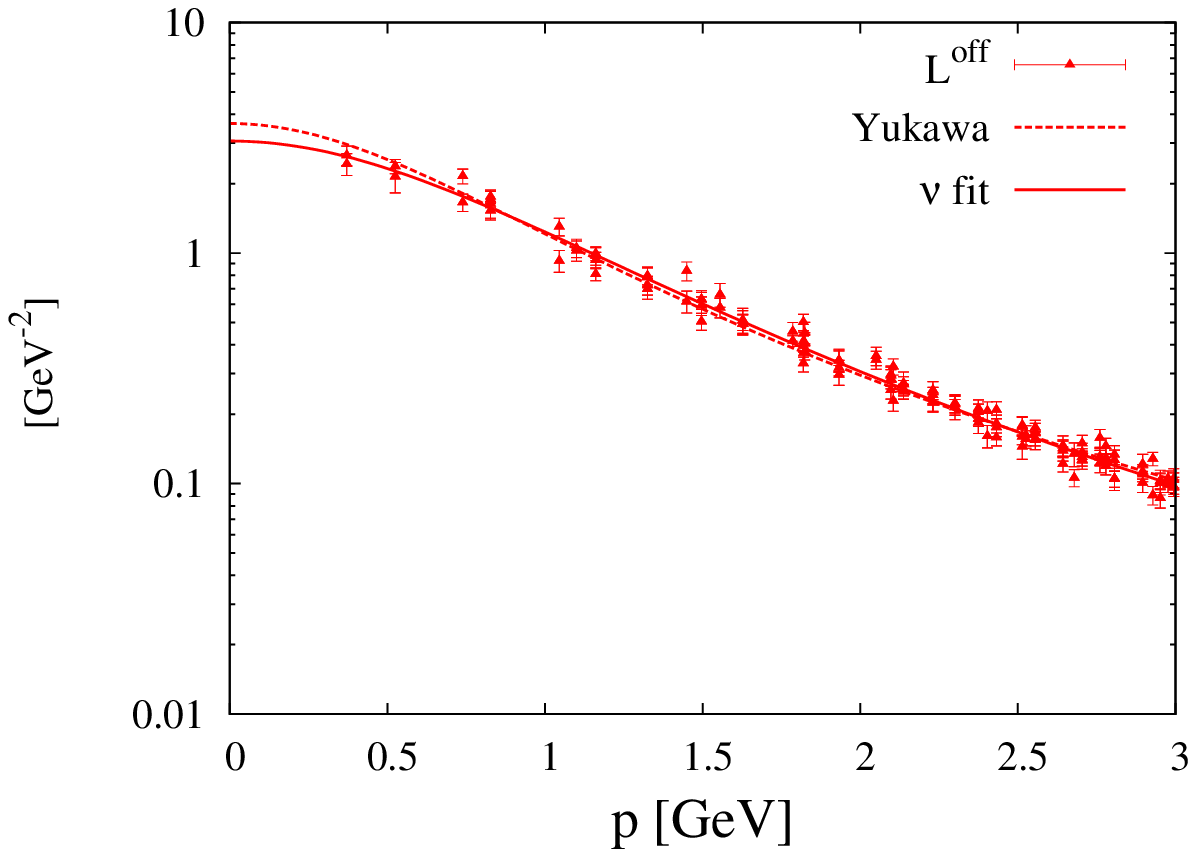}
\includegraphics[scale=0.5]{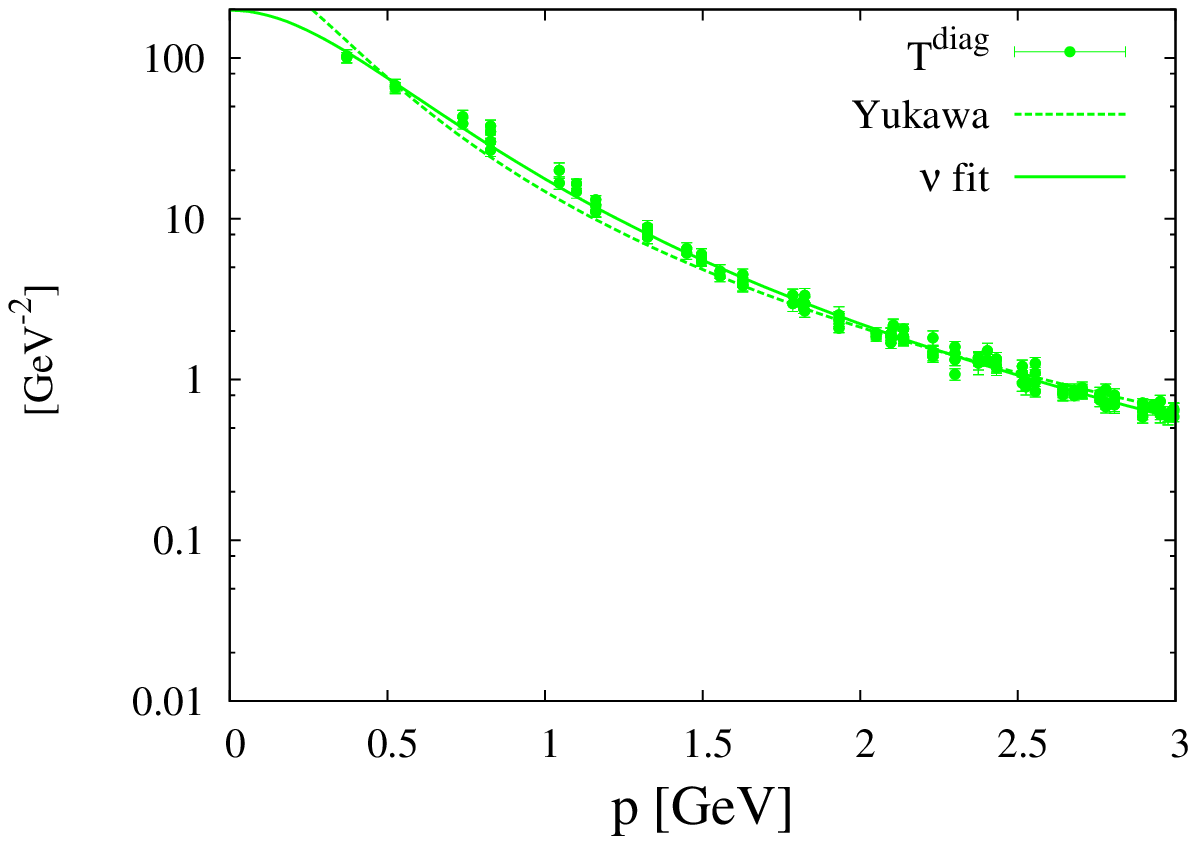}
\caption{
The logarithmic plot of $T^{\rm off} (p^2)$, $L^{\rm off} (p^2)$ and $T^{\rm diag} (p^2)$ as the function of the momentum $p\equiv (p_\mu p_\mu)^{1/2}$ with $32^4$ at $\beta$=6.0. The transverse and longitudinal parts in the off-diagonal gluon propagator are well described with four-dimensional Yukawa function, while the diagonal part is not well described in the infrared region. All parts of gluon propagators are well described with $\nu$-Ansatz (\ref{nu-fit}) in the whole region of $p<3.0 {\rm GeV}$.}
\label{Fig3}
\end{center}
\end{figure}

We show the fit results in Fig.\ref{Fig3} and the parameters in Table~\ref{TableIII}. As for the off-diagonal gluon propagator, both transverse part and longitudinal part are well described by the four-dimensional Euclidean Yukawa function, although, these parameters are a bit different. This result coincides with our previous work \cite{GIS12} and this means that the spectral function of off-diagonal gluons has a negative region \cite{GIS12,IS09}. On the other hand, the diagonal gluon propagator is not well described with the four-dimensional Euclidean Yukawa function.
\begin{table}[h]
\caption{ The fit results of the off-diagonal parts and the diagonal part by the four-dimensional Euclidean Yukawa function for $p<3.0 {\rm GeV}$. The transverse and longitudinal parts in the off-diagonal gluon propagator are well described by the four-dimensional Yukawa function with $\chi ^2 /N \simeq 1.4$, while the diagonal gluon propagator is not well described from the large $\chi ^2 /N$ ($\sim 2.4$).
}
\label{TableIII}
\begin{center} 
\begin{tabular}{cccccc}
\hline
\hline
&  lattice size ~  & $p_{\rm max} [\rm GeV]$     &   $m [{\rm GeV}] $  & $Z$ &   $\chi ^2/N$ \\
\hline
$T^{\rm off}$&$32^4$&    ~~~3.0~~~  & 1.27(0) &  6.08(4) & 1.4\\
$L^{\rm off}$&$32^4$&    ~~~3.0~~~  & 0.95(1) &  3.22(3) & 1.4\\
$T^{\rm diag}$&$32^4$&    ~~~3.0~~~ &  0.36(1)  & 17.7(1) & 2.4\\
\hline
\hline
\end{tabular}
\end{center} 
\end{table}

More generally, we fit the propagators by more general Ansatz,
\begin{eqnarray}
\frac{Z}{\left( p^2+m^2\right)^{\nu}} \label{nu-fit}
\end{eqnarray}
with the fit parameters $Z,$ $m$ and $\nu$ in the region of $p < p_{\rm max} = 3.0 {\rm GeV}$.
We call this ``$\nu$-Ansatz".
We show the fit results in Fig.\ref{Fig3} and the parameters in Table~\ref{TableIV}. The transverse and longitudinal parts of the off-diagonal propagator have different functional forms: The transverse part has $\nu \simeq 1.3-1.4$ and the mass parameter $m \simeq 1 {\rm GeV}$ and the longitudinal part has $\nu \simeq 1.8$ and $m \simeq 1 {\rm GeV}$. From this result, the transverse part behaves like a four-dimensional Yukawa function with the mass parameter $m \simeq 1 {\rm GeV}$. On the other hand, the longitudinal part does not behave like it ($\nu \simeq 1.8 > 1.5$), however, the mass parameter takes a similar value, $m \simeq 1 {\rm GeV}$. Due to this larger $\nu$, the longitudinal part is larger reduced than the transverse one in high momentum.
The diagonal part has $\nu \simeq 1.8$ and the mass parameter $0.6 {\rm GeV}$. The diagonal part is better described with these parameters than the four-dimensional Yukawa function. 
Note that both diagonal and off-diagonal gluon propagators show 
$\nu \ne 1$, which means that they cannot be described 
by a simple massive propagator with a definite pole mass.
\begin{table}[h]
\caption{ The fit results of the off-diagonal parts and the diagonal part by the $\nu$-Ansatz (\ref{nu-fit}) for $p<3.0 {\rm GeV}$. All these parts in gluon propagators are well described by this fit function with corresponding parameters.
}
\label{TableIV}
\begin{center} 
\begin{tabular}{ccccccc}
\hline
\hline
&  lattice size ~  & $p_{\rm max} [\rm GeV]$    &   $m [{\rm GeV}] $  &$\nu$ & $Z$ &   $\chi ^2/N$ \\
\hline
$T^{\rm off}$&$32^4$&    ~~~3.0~~~  & 1.09(2) &  1.34(2)&4.0(2) & 1.2\\
$L^{\rm off}$&$32^4$&    ~~~3.0~~~  & 1.21(4) &  1.75(4)&6.0(7)& 1.2\\
$T^{\rm diag}$&$32^4$&    ~~~3.0~~~ &  0.58(1)  & 1.76(1)&29(1) & 1.2\\
\hline
\hline
\end{tabular}
\end{center} 
\end{table}

\subsection{Spectral function of gluons in MA gauge}
The spectral function $\rho (\omega)$ is given by the inverse Laplace transformation of the zero-spatial-momentum propagator $D_0(t)$. Then we define
\begin{eqnarray}
\rho^{\mu \nu}(\omega) = \frac{1}{2\pi i} \int ^{c+i\infty}_{c-i\infty} dt e^{\omega t} D_0^{\mu\nu} (t),
\end{eqnarray}
where $D_0^{\mu \nu}(t)$ is defined with the gluon propagator $G^{\mu\nu}(x)$ by
\begin{eqnarray}
D_0^{\mu \nu}(t) \equiv \int d^3\vec{x} G^{\mu\nu}(\vec{x},t).
\end{eqnarray}
Note that $D_0^{\mu \nu}(t)$ and $\rho^{\mu \nu}(\omega)$ no longer have Lorentz tensor structure due to the spatial integral. 
Corresponding to the two components of the gluon propagator, 
\begin{eqnarray}
&~& G^{\mu\nu}(x) =\nonumber \\
&~&\int \frac{d^4p}{(2\pi) ^4  }e^{ip\cdot x}\left\{\left( \delta ^{\mu\nu} -\frac{p^\mu p^\nu}{p^2} \right) T(p^2) + \frac{p^\mu p^\nu}{p^2} L(p^2) \right\},~~~~
\end{eqnarray}
the zero-spatial-momentum propagator has also two components,
\begin{eqnarray}
D_0^{\mu \nu}(t) &=& \left( \delta ^{\mu\nu} -\delta ^{\mu 4}\delta ^{\nu 4}\right)  \int \frac{dp_4}{(2\pi)}e^{ip_4 t} T(p_4^2) \nonumber \\
&+& \delta ^{\mu 4}\delta ^{\nu 4} \int \frac{dp_4}{(2\pi)}e^{ip_4 t} L(p_4^2)\nonumber\\
&\equiv&\left( \delta ^{\mu\nu} -\delta ^{\mu 4}\delta ^{\nu 4}\right)  D_T(t)+
 \delta ^{\mu 4}\delta ^{\nu 4} D_L(t).
\end{eqnarray}
Note that $D_T(t)$ is obtained from the space-space correlator and $D_L(t)$ is from the time-time correlator.
Thus, we can define the two components of the spectral function as
\begin{eqnarray}
\rho^{\mu \nu}(\omega) = \left( \delta ^{\mu\nu} -\delta ^{\mu 4}\delta ^{\nu 4}\right)\rho _T(\omega) + \delta ^{\mu 4}\delta ^{\nu 4} \rho _L(\omega),
\end{eqnarray}
where
\begin{eqnarray}
\rho _{T,L}(\omega) =\frac{1}{2\pi i} \int ^{c+i\infty}_{c-i\infty} dt e^{\omega t}D_{T,L}(t) .
\end{eqnarray}
$\rho _{T,L}(\omega)$ denotes the spectral function of transverse and longitudinal gluons.
In the same way, we can define the corresponding effective mass,
\begin{eqnarray}
M_{T,L} (t)\equiv \ln \frac{D_{T,L} (t)}{D_{T,L} (t+1)} 
\end{eqnarray}
In the Landau gauge, as $t$ increases, the effective mass increases. If the spectral function $\rho (\omega)$ is non-negative, the effective mass $M(t)$ should be a monotonically decreasing function. In fact, if $M(t)$ has an increasing part, $\rho(\omega)$ should have negative region \cite{MO87}.

If the functional form of some part (longitudinal or transverse part) of the gluon propagators is well described by the  $\nu$-Ansatz, $\frac{Z}{(p^2+m^2)^\nu}~ \left(\nu >0\right)$, the corresponding zero-spatial-momentum propagator $D_\nu (t)$ is obtained by
\begin{eqnarray}
D_\nu (t) &\equiv &\int \frac{dp_4}{(2\pi)}e^{ip_4 t}\frac{Z}{\left(p_4^2+m^2\right)^{\nu }} \nonumber \\
&=& \frac{Z}{\sqrt{\pi}\Gamma \left(\nu \right)}\left(\frac{t}{2m}\right)^{\nu-1/2} K_{\nu -1/2} (mt) ~~ \left({\rm Re} t >0 \right), ~~~~~
\end{eqnarray} 
where $K_{\nu -1/2} $ is the modified Bessel function.
Therefore, the corresponding effective mass $M_\nu (t)$ is expressed by
\begin{eqnarray}
M_\nu (t) &\equiv& \ln \frac{D_\nu (t)}{D_\nu (t+1)} \nonumber\\
&=&\ln \frac{t^{\nu -1/2}K_{\nu -1/2} (mt) }{\left( t+1\right)^{\nu -1/2}K_{\nu -1/2} \left( m\left( t+1\right) \right)}. \label{theoritical curve}
\end{eqnarray}
The modified Bessel function is reduced to $K_{\nu -1/2}(mt) \sim (mt)^{-1/2}e^{-mt}$ for large $mt$, and thus  
\begin{eqnarray}
M_\nu (t) &\simeq &\ln\frac{t^{\nu -1}}{(t+1)^{\nu -1}e^{-m}} \nonumber \\
&\simeq &m - \left( \nu - 1 \right) \frac{1}{t}.
\end{eqnarray}
This indicates that, if the functional form of the propagator is well described by the $\nu$-Ansatz (\ref{nu-fit}), $M_\nu (t)$ increases for $\nu >1$ as $t$ increases. Due to this, the spectral function would have a negative region.

\begin{figure}[h]
\begin{center}
\includegraphics[scale=0.5]{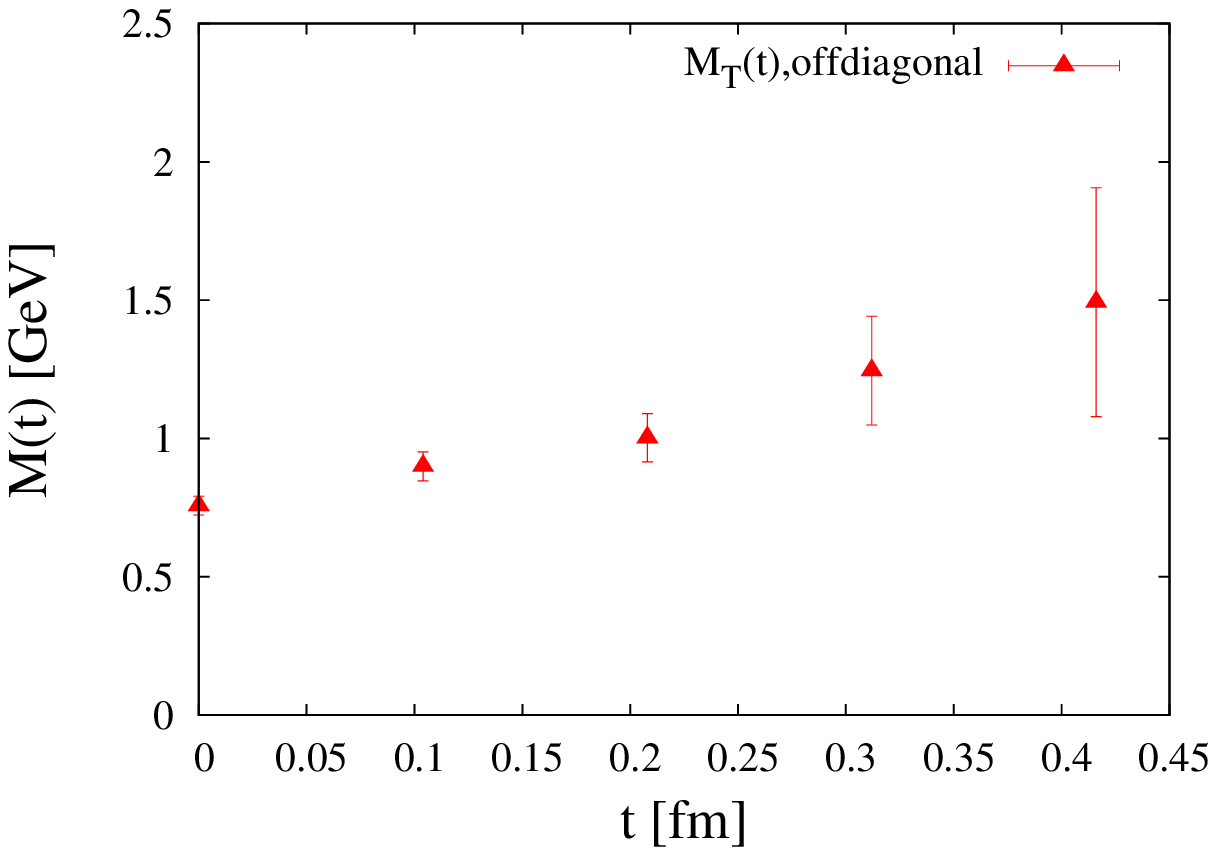}
\includegraphics[scale=0.5]{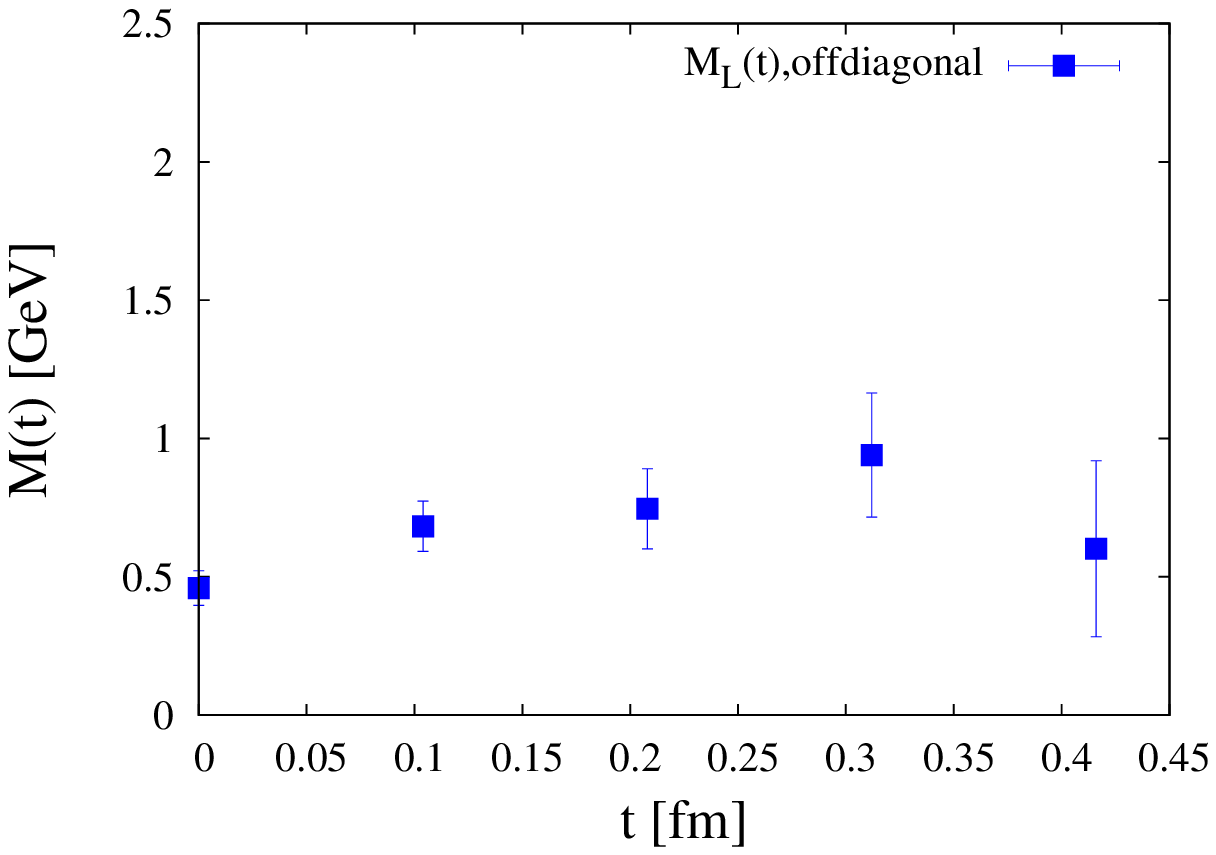}
\includegraphics[scale=0.5]{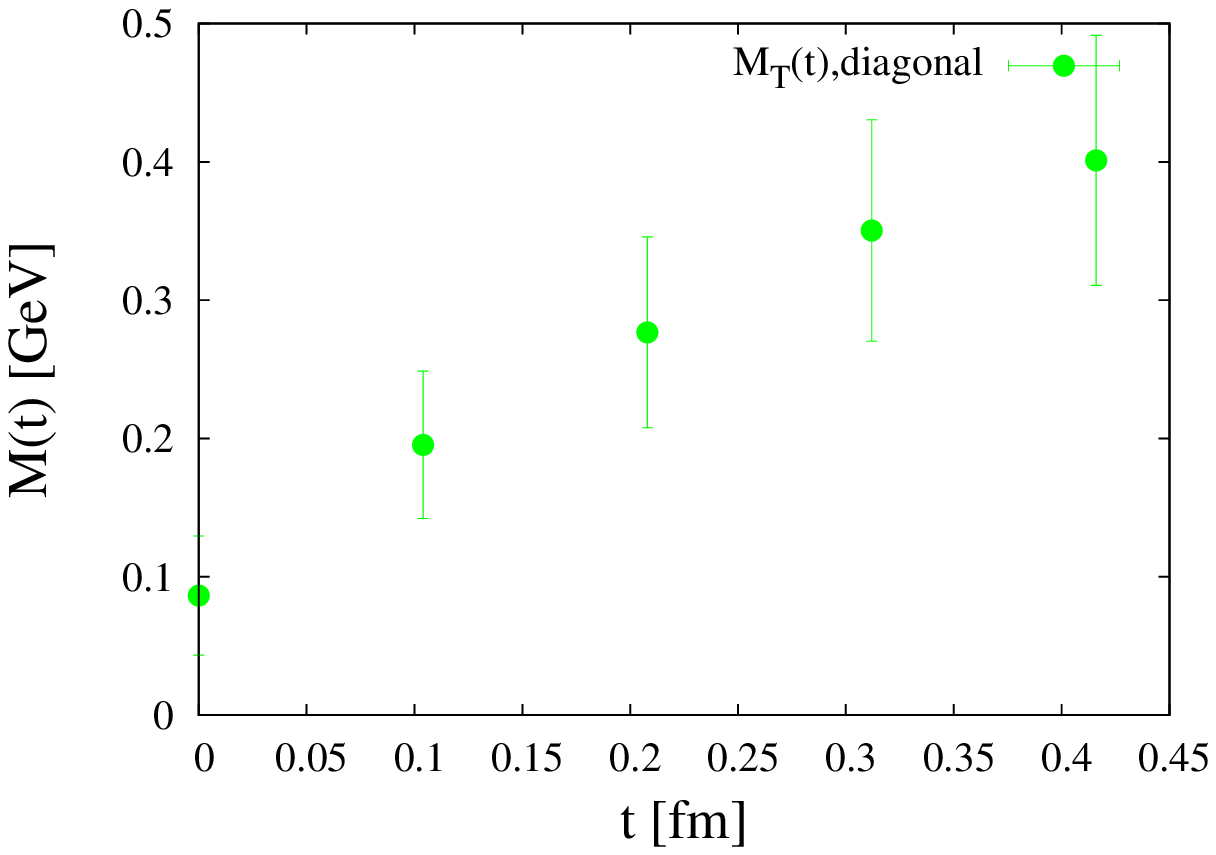}
\caption{
The effective mass plot of the longitudinal and transverse parts of off-diagonal gluons, $M_L(t)$ and $M_T(t)$, and the diagonal gluon. The off-diagonal effective masses are obtained on the $16^4$ lattice with $\beta =6.0$ using 200 configurations, and the diagonal effective mass is obtained on the $32^4$ lattice with $\beta =6.0$ using 40 configurations. Each effective mass $M(t)$ seems to be an increasing function of $t$.}
\label{Fig3d-e}
\end{center}
\end{figure}

In fact, we calculate the off-diagonal effective masses and the diagonal effective mass. There are two off-diagonal effective masses corresponding to the longitudinal and transverse parts, however, there is only the transverse part in the diagonal effective mass, due to $\partial _\mu A_\mu ^a \simeq 0~ (a=3,8)$.
In Fig.\ref{Fig3d-e}, we show these effective masses in SU(3) MA gauge with U(1)$_3\times$ U(1)$_8$ Landau gauge. We calculate the off-diagonal effective masses on $16^4$ with $\beta = 6.0$ and the diagonal effective mass on $32^4$ with $\beta =6.0$ due to the large dependence of the lattice size. For the diagonal gluon and the off-diagonal transverse/longitudinal
gluons, each effective mass $M(t)$ is found to be
an increasing function of $t$ at least in small $t$ region,
which leads to existence of negative region of
the spectral function of each gluon component \cite{MO87}.

\section{Summary and Concluding Remarks}
\label{6}
We have studied the gluon propagators in the MA gauge with the U(1)$_3\times$U(1)$_8$ Landau gauge fixing using the SU(3) lattice QCD both in coordinate space and in momentum space. The Monte Carlo simulation is performed on the $16^4$ lattice with $\beta$ = 6.0 and $32^4$ with $\beta$ =5.8 and 6.0 at the quenched level.

In coordinate space, we have calculated the Euclidean scalar combinations of the propagators $G_{\mu\mu} (r)$ in the diagonal and the off-diagonal gluons and estimated each gluon mass from the linear slope of the logarithmic plot of $r^{3/2}G_{\mu\mu}(r)$
We have found that 
the off-diagonal gluons behave as massive vector bosons with the effective mass $M_{\rm off} \simeq 1$ GeV for $r =0.4 -1.0$ fm, while the diagonal gluons behave as lighter vector bosons with $M_{\rm diag} \simeq 0.3$ GeV for $r =0.4 -1.0$ fm.  Due to the larger off-diagonal gluon mass, the off-diagonal gluons cannot mediate the interaction over the large distance as $r \gg M_{\rm off}^{-1}$, while the diagonal gluons can propagate in this region. Such a behavior would lead infrared Abelian dominance in the MA gauge.

Furthermore, we have investigated the diagonal gluon propagator and off-diagonal gluon propagator in momentum space. To our knowledge, the analysis of the gluon propagators with the MA gauge of SU(3) lattice in momentum space is the first study. In the MA gauge with U(1)$_3\times$ U(1)$_8$ Landau gauge, the diagonal propagator has the only one component and the off-diagonal propagator has two components, the transverse part and the longitudinal part. In the infrared region, the diagonal propagator is enhanced, while the transverse and the longitudinal parts in the off-diagonal propagator show the similar suppressed behavior. In this way, the infrared Abelian dominance is also found in momentum space as in coordinate space.
Furthermore, at zero momentum, the two components of the off-diagonal propagator are finite and relatively suppressed and the diagonal propagator seems to be finite and enhanced. However, as for the diagonal propagator in the deep infrared region, more careful analysis with a larger lattice size would be required for the definite argument. 

We have also estimated these gluon masses from these components of diagonal propagator and off-diagonal propagator in momentum space by compared with the Proca formalism in the region of $p < 1.1 {\rm GeV}$ as in coordinate space. As a result, the off-diagonal gluon mass seems to be $M_{\rm off} \simeq 1$ GeV and the diagonal gluon mass seems to be $M_{\rm diag} \simeq 0.3$ GeV. These results are consistent with the analysis of coordinate space.

In addition, we have also investigated the functional form of the propagator in the MA gauge. The gluon propagators have been fitted by $Z/\left(p^2+m^2\right)^{\nu}$ with the parameters, $Z, m$ and $\nu$, in the region of $p<3.0{\rm GeV}$. These propagators are well described with the functional form. The best fit results show that these mass parameters $m_{\rm diag}, m^T_{\rm off}$ and $m^L_{\rm off}$ have a relation, $m^L_{\rm off},m^T_{\rm off} \gg m_{\rm diag}$ and all of these exponentiation parameters, $\nu$, are larger than unity. Thus even the functional form of the diagonal propagator is larger enhanced than that of the off-diagonal propagator and would be found infrared Abelian dominance. Furthermore, from these fit results, in particular $\nu >1$, all of the corresponding spectral functions would have negative regions as expected in our previous work \cite{GIS12}. In fact, we have calculated these zero-spatial-momentum propagators, $D_0(t)$, and  the corresponding effective mass, $M(t) \equiv  \ln D_0(t)/D_0 (t+1)$. We expect the negative regions of these spectral functions from the behaviors of these effective mass. 

\section*{Acknowledgements}
The authors are deeply grateful to Dr. H. Iida and Dr. T. Z. Nakano for useful discussions.  
This work is supported in part by the Grant for Scientific Research 
[(C) No.~23540306, Priority Areas ``New Hadrons'' (E01:21105006)], Grant-in-Aid for JSPS Fellows (No.24-1458)
from the Ministry of Education, Culture, Science and Technology 
(MEXT) of Japan, and the Global COE Program, 
``The Next Generation of Physics, Spun from Universality and Emergence".
The lattice QCD calculations are done on NEC SX-8R at Osaka University.

\appendix
\section{Comparison of Cartan decomposition and ordinary decomposition}
\label{ap1}

In SU($N_c$) lattice QCD, the gluon field $A_\mu$ is usually 
defined from the link-variable $U_\mu$ based on \cite{R12}
\begin{eqnarray}
 U_\mu (x) =  e^{iag A_\mu (x)}\in {\rm SU}(N_c),
\end{eqnarray}
which we call ordinary decomposition (OD).
In our paper, this gluon field $A_\mu$ 
is mainly used for the argument of gluon propagators.
However, in terms of the partial gauge fixing in the MA gauge, 
one can take other definition of the gluon field ${\cal A}_\mu$ 
based on the Cartan decomposition (CD), 
\begin{eqnarray}
U_\mu (x) &=& e^{iag  \sum_{a \neq 3,8}{\cal A}_\mu ^a(x) T^a}
\cdot e^{iag  \sum_{a = 3,8}{\cal A}_\mu ^a(x) T^a} \nonumber \\ 
&\in& {\rm SU}(N_c)/{\rm U(1)}^{N_c-1}\times {\rm U(1)}^{N_c-1}. 
\end{eqnarray}
In the continuum limit, these two definitions of gluon fields coincide.
Actually, in our lattice calculation, 
even with these two definitions of gluons, 
we obtain almost the same results for both 
diagonal and off-diagonal propagators. 

In this appendix, we compare 
the diagonal gluon propagators with OD and CD, 
which are defined by
\begin{eqnarray}
&~&\frac{1}{2} \sum_{a= 3,8} \left< A_\mu^a(x)A_\nu^a(y)\right>~~\left({\rm OD}\right), \nonumber \\
&~&\frac{1}{2} \sum_{a= 3,8} \left< {\cal A}_\mu^a(x){\cal A}_\nu^a(y)\right>~~\left({\rm CD}\right).
\end{eqnarray}

\begin{figure}[h]
\begin{center}
\includegraphics[scale=0.5]{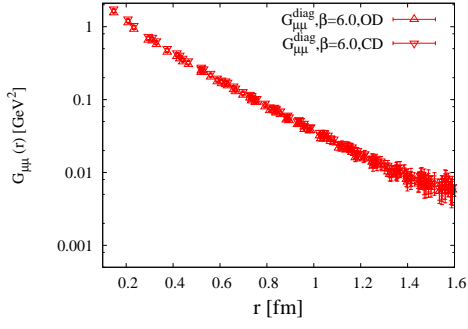}
\caption{ The SU(3) lattice QCD results of $G_{\mu\mu}^{\rm diag}(r)$ with Cartan decomposition (CD) and ordinary decomposition (OD) in the MA gauge with the U(1)$_3\times$U(1)$_8$ Landau gauge fixing. The Monte Carlo simulation is performed on the $32^4$ lattice with $\beta$ = 6.0. The CD gluon propagator almost coincides with the OD gluon propagator.}
\label{Fig5a}
\end{center}
\end{figure}

In Fig.\ref{Fig5a}, we show the logarithmic plot of the scalar combination of the diagonal gluon propagators with OD and CD in coordinate space. This result is obtained on $32^4$ with $\beta = 6.0$. These propagators are almost the same behavior. 

\begin{figure}[h]
\begin{center}
\includegraphics[scale=0.4]{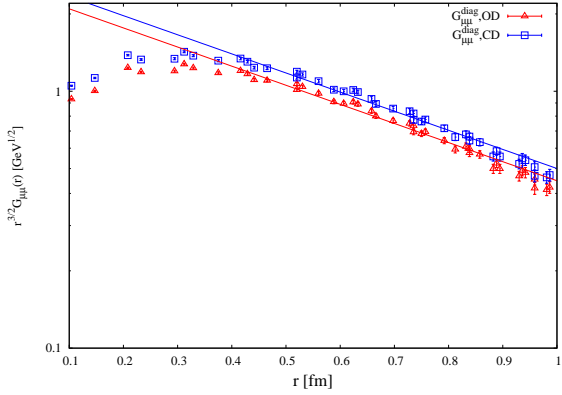}
\caption{ The logarithmic plot of $r^{3/2}G_{\mu\mu}^{\rm diag}(r)$ with Cartan decomposition (CD) and ordinary decomposition (OD) in the MA gauge with the U(1)$_3\times$U(1)$_8$ Landau gauge fixing on the $32^4$ lattice at $\beta$=6.0. The slopes of these plots in the region of $r=0.4-1.0 {\rm fm}$ are almost the same.}
\label{Fig5b}
\end{center}
\end{figure}

Furthermore, we estimate each diagonal gluon mass from the slope on the logarithmic plot of 
$r^{3/2}G_{\mu\mu}^{\rm diag} (r)$ as in Sec.\ref{4}. In Fig.\ref{Fig5b}, we show the logarithmic plot of  $r^{3/2}G_{\mu\mu}^{\rm diag} (r)$ with OD and CD. These gluon masses are obtained from the slope analysis in the range of $r =0.4-0.8 {\rm fm}$. In both cases, diagonal gluon mass seems to be about 0.3 ${\rm GeV}$. Also for the off-diagonal gluons, the CD results are almost the same as the OD results.

\section{The longitudinal part of the diagonal propagator}
The lattice global Landau-gauge fixing is given by maximization of
\begin{eqnarray}
R[U]= {\rm ReTr}\sum _{\mu , x} U_\mu (x)
\end{eqnarray}
From this global gauge condition, the local gauge fixing is derived as
\begin{eqnarray}
\Delta R &\equiv& R[U^V] -R[U^{V=1}] \nonumber \\ 
&=&{\rm ReTr}\sum _{\mu } [VU_\mu (x) + U_\mu (x-\hat{\mu})V^{\dagger} -(V=1)] \nonumber \\
&\leq & 0,
\end{eqnarray}
where $V\in SU(3)$ denotes the arbitrary local gauge transformation on the site $x$.
By expanding the local gauge transformation as $V\equiv e^{i\theta ^a T^a}\simeq 1+i\theta ^a T^a$, the local gauge formation is obtained,
\begin{eqnarray}
&{\rm Tr}&T^a \sum_\mu \left\{U_\mu (x) - U_\mu ^\dagger (x) - \left( U_\mu  (x-\hat{\mu})-U_\mu ^\dagger (x-\hat{\mu}) \right)\right\} \nonumber \\
&=&0.
\end{eqnarray}
In the Landau gauge, the gluons are usually defined by
\begin{eqnarray}
A_\mu (x) \equiv \frac{1}{2i} \left\{ U_\mu (x) -U_\mu ^\dagger (x) \right\} -\left( \rm trace~part \right) .
\end{eqnarray}
With this definition, the local gauge fixing is expressed by the ordinary Landau gauge fixing,
\begin{eqnarray}
{\rm Tr}&T^a\sum _\mu \left\{ A_\mu (x)-A_\mu (x-\hat{\mu}) \right\}= 0.
\end{eqnarray}
However, with our definition of gluons Eq. (\ref{gluon definition}), the local gauge fixing condition is given by
\begin{eqnarray}
{\rm Tr}&T^a\sum _\mu \left\{ \sin A_\mu (x)-\sin A_\mu (x-\hat{\mu}) \right\}= 0,
\end{eqnarray}
where we use precise relation of $\sin A_\mu (x)  \equiv \frac{1}{2i}\left\{ U_\mu (x) -U_\mu ^\dagger (x) \right\}$. In small lattice spacing, this local gauge fixing corresponds to the ordinary Landau gauge.
In general, however, this gauge fixing condition does not coincide with the ordinary gauge fixing condition, and thus, the gluon propagator with our definition of gluons in momentum space would have the longitudinal part as well as the transverse part.

The similar situation arises for diagonal gluons in MA gauge with U(1)$_3 \times$ U(1)$_8$ Landau gauge fixing, because the gauge fixing for U(1)$_3 \times$ U(1)$_8$ part is given by the maximization of 
\begin{eqnarray}
R_{\rm U(1)L}\equiv \sum_x \sum_{\mu=1}^4 {\rm Re}\ {\rm tr}[u_\mu(x)].
											\label{eqn:defU1Landau}
\end{eqnarray}
Thus, the local condition is given by
\begin{eqnarray}
\sum _\mu \left\{ \sin {\cal A}_\mu^a (x)-\sin {\cal A}_\mu^a (x-\hat{\mu}) \right\}= 0~~(a=3,8), \label{lattice landau}
\end{eqnarray}
where $\vec{\cal A}_\mu (x) =({\cal A}_\mu ^3(x),{\cal A}_\mu^8 (x))$ is defined by $u_\mu (x)=e^{iag  \sum_{a = 3,8}{\cal A}_\mu ^a(x) T^a}$. This condition (\ref{lattice landau}) is satisfied accurately in numerical calculation.
In this gauge, even with small lattice spacing, it is not so trivial whether the local gauge condition (\ref{lattice landau}) coincides with the ordinary Landau gauge due to the large fluctuation of diagonal gluons. Note that there are two different definitions of diagonal gluons, ${\cal A} _\mu ^a(x)$ and $A_\mu ^a (x)$, however, the behavior of these gluon propagators would be similar as shown in Appendix \ref{ap1}.

In Fig.\ref{Fig6}, we show the longitudinal part and the transverse part of the diagonal propagators. The transverse part is dominated in whole momentum region,
\begin{eqnarray}
L^{\rm diag} (p^2)\ll T^{\rm diag} (p^2).
\end{eqnarray}
 Therefore, there would be the only transverse part in the diagonal gluon propagator even with our definition of gluons,
 \begin{eqnarray}
G_{\mu\nu}^{\rm diag}(p) \simeq \left(\delta _{\mu\nu} - \frac{p_\mu p_\nu}{p^2}\right) T^{\rm diag}(p^2).
\end{eqnarray} 

Note that as for the off-diagonal gluon propagator, there are both longitudinal and transverse parts even in the continuum limit,
\begin{eqnarray}
G_{\mu\nu}^{\rm off}(p) = \left(\delta _{\mu\nu} - \frac{p_\mu p_\nu}{p^2}\right) T^{\rm off}(p^2) + \frac{p_\mu p_\nu}{p^2}L^{\rm off}(p^2).~~
\end{eqnarray}

\label{ap2}
\begin{figure}[h]
\begin{center}
\includegraphics[scale=0.5]{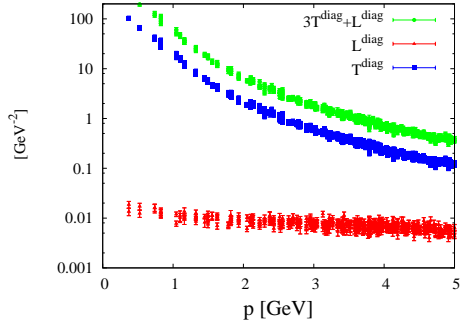}
\caption{
The logarithmic plot of $T^{\rm diag} (p^2)$, $L^{\rm diag} (p^2)$ and $G_{\mu\mu}^{\rm diag}(p^2)=3T^{\rm diag} (p^2)+L^{\rm diag} (p^2)$ as the function of the momentum $p\equiv (p_\mu p_\mu)^{1/2}$ in the MA gauge with U(1)$_3\times$U(1)$_8$ Landau gauge fixing with $32^4$ at $\beta$=6.0. In this gauge, $L^{\rm diag} (p^2)\ll T^{\rm diag} (p^2)$ is satisfied at least in the region of $p<3.0 {\rm GeV}$}.
\label{Fig6}
\end{center}
\end{figure}

\end{document}